\documentclass{phb-proc4-auth}

\usepackage{graphicx}
\usepackage{amssymb}
\usepackage{epsfig}
\usepackage{bm}

\begin{document}
\begin{frontmatter}

\journal{SCES '04}

\title{Low energy spin fluctuations in the heavy fermion compound
Ce$_{0.925}$La$_{0.075}$Ru$_{2}$Si$_{2}$}

\author[a]{W. Knafo}
\author[a]{S. Raymond\corauthref{1}}
\author[a]{B. F{\aa}k}
\author[b]{M.A. Adams}
\author[c]{P. Lejay}
\author[a]{J. Flouquet}

\address[a]{CEA-Grenoble, DSM/DRFMC/SPSMS, 38054 Grenoble
Cedex 9, France}
\address[b]{ISIS Facility, Rutherford Appleton
Laboratory, Chilton, Didcot,
 Oxon OX11 0QX, UK}
\address[c]{CRTBT, CNRS, B.P. 166, 38042 Grenoble Cedex 9, France}

\corauth[1]{Corresponding Author. Phone: +33(0)438783738,
 Fax: +33(0)438785109, Email: raymond@drfmc.ceng.cea.fr}

\begin{abstract}

We report inelastic neutron scattering measurements performed on a
single crystal of the heavy fermion compound
Ce$_{0.925}$La$_{0.075}$Ru$_{2}$Si$_{2}$, which is at the
borderline between an antiferromagnetically ordered and a
paramagnetic ground state. Intensity maps as a function of
wavevector and energy ($0.1<E<1.2$ meV) were obtained at
temperatures $T=0.1$ and 2 K, using the time-of-flight
spectrometer IRIS. An unexpected saturation of the relaxation rate
and static susceptibility of the spin fluctuations is found at low
temperatures.

\end{abstract}

\begin{keyword}

Heavy Fermions \sep Quantum Critical Point \sep Spin Fluctuations
\sep Inelastic Neutron Scattering

\end{keyword}

\end{frontmatter}

The heavy fermion system Ce$_{1-x}$La$_{x}$Ru$_{2}$Si$_{2}$ is
characterized by a quantum phase transition that occurs at $T=0$
for the critical concentration $x_{c}=7.5\%$ \cite{Raymond2001}.
The ground state of this 3D Ising system is a non magnetically
ordered Fermi liquid for $x<x_{c}$ and is antiferromagnetically
ordered with the propagation vector $\mathbf{k}_{1}=$(0.31 0 0)
for $x>x_{c}$. While a long range order is only established for
$x>x_{c}$, a short range magnetic order corresponding to a small
moment of 0.02 $\mu_{B}$ develops below 2 K at $x_{c}$. Both parts
of this phase diagram are characterized by the same kind of low
energy spin fluctuations: a broadened quasielastic signal ascribed
to Kondo effect is obtained for any wavevector $\mathbf{k}$
sufficiently far from the incommensurate wavevectors
$\mathbf{k}_{1}$, $\mathbf{k}_{2}=(0.31,0.31,0)$, and
$\mathbf{k}_{3}=(0,0,0.35)$ (and equivalents), at which short
range RKKY interactions lead to an enhancement of the signal
\cite{Raymond97,Knafo,Kadowaki2}.

We report here an inelastic neutron scattering study of the
compound of critical concentration $x_{c}$, for which very low
temperatures and energy transfers were obtained using a dilution
refrigerator on the inverted-geometry time-of-flight spectrometer
IRIS, at the neutron spallation source of ISIS (Didcot, U.K.). A
single crystal of 250 mm$^{3}$ was grown by the Czochralski
method. Ce$_{0.925}$La$_{0.075}$Ru$_{2}$Si$_{2}$ crystallizes in
the body-centered tetragonal I4/mmm space group with lattice
parameters $a = 4.197$ $\rm{\AA}$ and $c=9.797$ \AA. The
experiment was carried out using a fixed final neutron energy of
1.84 meV (PG (002) analyzer) resulting in a resolution of 18
$\mu$eV (FWHM of the incoherent elastic peak). The tail of the
elastic peak limits the study of excitations to a range of
accessible energy transfers $E$ higher than 0.1 meV. Measurements
have been performed at 100 mK and 2 K in the $(0,0,1)$ plane for
three orientations a), b), and c) of the crystal, which correspond
to the angles $\psi_{a}=114.98^\circ$, $\psi_{b}=70.5^\circ$, and
$\psi_{c}=75.74^\circ$ between the direction $[1,0,0]$ of the
crystal and the incident neutron beam.

\begin{figure}[h]
    \centering
    \epsfig{file=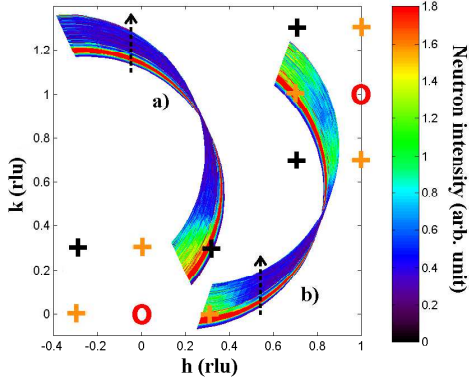,height=50mm}
    \caption{Mappings of the excitations measured at $T=0.1$ K for the configurations a) and b).
    The red circles represent the positions $(h k 0)$ of the nuclear Bragg peaks
    and the orange and black crosses represents the
    wavevectors $\mathbf{k_{1}}$ and $\mathbf{k_{2}}$,where magnetic correlations are observed, respectively. For each configuration,
    the dashed arrow indicates the direction of increasing values of $E$ from -0.2 to 1.2 meV, the red
    band corresponding to the incoherent elastic scattering obtained at $E=0$ (color online).}
\end{figure}

Figure 1 shows the $(\mathbf{Q},E)$ intensity maps of the
excitations obtained in configurations a) and b) at $T=100$ mK.
Each point of this map corresponds to a momentum transfer
$\mathbf{Q}=(h,k,0)$ associated with an energy transfer $E$.
Constant-$E$ lines are circular trajectories in reciprocal space,
the $E=0$ lines corresponding to the red bands of Fig. 1. The
orientations a), b), and c) of the crystal have been chosen so
that the $E=0$ line crosses a momentum transfer
$\mathbf{Q}={\bm\tau}+\mathbf{k}$, where ${\bm\tau}=(h,k,0)$ is a
nuclear Bragg peak, and where the wavevector $\mathbf{k}$
corresponds to $\mathbf{k}_{1}$ for configurations b) and c), and
$\mathbf{k}_{2}$ for configuration a). As already reported in a
previous study on a triple-axis spectrometer with a resolution of
150 $\mu$eV \cite{Raymond97}, a magnetic Bragg peak is obtained at
$T=0.1$ K at the wavevector $\mathbf{k}_{1}$, which is related to
a small moment short range order. This peak is absent at $T=2$ K.
An enhancement of spin fluctuations is observed in the vicinity of
the momentum transfers, which correspond to $\mathbf{k}_{1}$ or
$\mathbf{k}_{2}$. The signal is also found to be independent of
the wavevector if $\mathbf{k}$ is sufficiently far from
$\mathbf{k}_{1}$ and $\mathbf{k}_{2}$ (and also from
$\mathbf{k}_{3}$), which corresponds to an important part of
reciprocal space.

For the configuration c), the angular integration of the signal at
low scattering angles $20<2\theta<30^\circ$ gives a spectrum for
$\mathbf{Q}$ in the vicinity of $\mathbf{Q}_{1}=(0.31,0,0)$ that
corresponds to the antiferromagnetic spin fluctuations at
$\mathbf{k}\simeq\mathbf{k}_{1}$. For $50<2\theta<80^\circ$, there
are no magnetic correlations and a smaller signal is obtained
after integration. Those two signals are shown as function of
energy at $T=100$ mK in figure 2, together with the fit of the
low-angle spectrum. This fit is made using an elastic and a
quasielastic contribution, the peak at 0.25 meV being spurious and
considered as part of the background. The elastic contribution is
the sum of the incoherent scattering and the magnetic Bragg peak.
Knowing that the scattered intensity is proportional to
$S(\mathbf{Q},E,T)$, which is related to the imaginary part of the
dynamical susceptibility $\chi''(\mathbf{Q},E,T)$ by:
\begin{eqnarray}
  S(\mathbf{Q},E,T) &=&
  \frac{1}{\pi}\frac{1}{1-e^{-E/k_{B}T}}\chi''(\mathbf{Q},E,T),\nonumber
\end{eqnarray}
the spin fluctuations for $\mathbf{Q}$ in the vicinity of
$\mathbf{Q}_{1}$ are fitted using the quasielastic Lorentzian
shape:
\begin{eqnarray}
  \chi''(\mathbf{Q}_{1},E,T) =
  \chi'(\mathbf{Q}_{1},T)\frac{E/\Gamma(\mathbf{Q}_{1},T)}{1+(E/\Gamma(\mathbf{Q}_{1},T))^{2}}.\nonumber
\end{eqnarray}

$\chi''(\mathbf{Q}_{1},E,T)$ is found to be independent of
temperature for $T\leq2$ K, the same relaxation rate
$\Gamma(\mathbf{Q}_{1},T)=0.29\pm0.2$ meV and static
susceptibility $\chi'(\mathbf{Q}_{1},T)$ being obtained for both
temperatures $T=0.1$ and 2 K.

\begin{figure}[h]
    \centering
    \epsfig{file=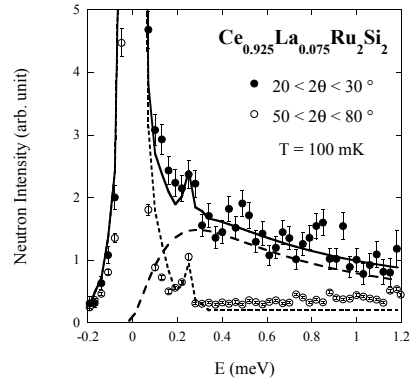,height=50mm}
    \caption{ Inelastic spectra obtained after angular integration of the signal in the range
    $20<2\theta<30^\circ$(filled circles) and for $50<2\theta<80^\circ$ (unfilled circles) at
    $T=0.1$ K in configuration c). The long dashed and short dashed lines show the
    quasielastic and elastic (plus background) contribution, respectively, and the solid line
    shows the total fit of the low angle spectrum.}
\end{figure}

This experiment, which benefited from the high energy resolution
of the time-of-flight spectrometer IRIS, confirms thus the static
moment previously measured using a triple-axis spectrometer
\cite{Raymond97}, and establishes clearly the saturation of the
relaxation rate $\Gamma(\mathbf{Q}_{1},T)$ and of the static
susceptibility $\chi'(\mathbf{Q}_{1},T)$ at low temperatures,
reported in \cite{Knafo}. This last result is unexpected, since
theories on second order quantum phase transitions predict a
diverging static susceptibility $\chi'(\mathbf{Q}_{1},T)$ and a
relaxation rate $\Gamma(\mathbf{Q}_{1},T)\rightarrow0$ in the
limit $T\rightarrow0$ \cite{Millis}.

\end{document}